\documentclass[aps,prc,twocolumn,10pt, superscriptaddress, showpacs, floatfix]{revtex4-1}
\usepackage{amssymb,epsfig}

\graphicspath{{figures/}}
\newcommand{\be}{\begin{equation}}
\newcommand{\ee}{\end{equation}}
\newcommand{\bea}{\begin{eqnarray}}
\newcommand{\eea}{\end{eqnarray}}

\begin{document}

\title{Finite-temperature relativistic nuclear field theory: an application to the dipole response}
\author{Elena Litvinova}
\affiliation{Department of Physics, Western Michigan University, Kalamazoo, MI 49008, USA}
\affiliation{National Superconducting Cyclotron Laboratory, Michigan State University, East Lansing, MI 48824, USA}
\author{Herlik Wibowo}
\affiliation{Department of Physics, Western Michigan University, Kalamazoo, MI 49008, USA}

\date{\today}

\begin{abstract}
Nuclear response theory beyond the one-loop approximation is formulated for the case of finite temperature. For this purpose, the time blocking approximation to the time-dependent part of the in-medium nucleon-nucleon interaction amplitude is adopted for the thermal (imaginary-time) Green's function formalism. We found that introducing a soft blocking, instead of a sharp blocking at zero temperature, brings the Bethe-Salpeter equation to a single frequency variable equation also at finite temperatures.
The method is implemented self-consistently in the framework of Quantum Hadrodynamics and designed to connect the high-energy scale of heavy mesons and the low-energy domain of nuclear medium polarization effects in a parameter-free way.  In this framework, we investigate the temperature dependence of dipole spectra in the even-even nuclei $^{48}$Ca and $^{100,120,132}$Sn with a special focus on the giant dipole resonance's width problem and on the low-energy dipole strength distribution.
\end{abstract}

\maketitle

%===============================================================================
% Introduction
{\it Introduction. \textemdash} Behavior of nuclear systems at finite temperature remains among the most difficult problems in nuclear theory. 
Nuclear temperature is a conventional concept for describing highly excited compound nuclei in long-lived intermediate 
states of such processes as neutron or proton capture, fusion, fission and heavy ion collisions. An accurate description of the
response of the compound nucleus to external probes and of its deexcitations is, therefore, of great importance for many applications, in particular, for nuclear data, nuclear technology, and modeling of astrophysical processes including the evolution of stars and galaxies \cite{MumpowerSurmanMcLaughlinEtAl2016}. 

The dipole response, which dominates spectra of nuclear excitations, is commonly divided into the high-energy part represented by the broad giant dipole resonance (GDR) associated with a high-frequency collective oscillation of protons and neutrons against each other, and the low-energy sector, the pygmy dipole resonance (PDR), which represents the oscillation of the neutron excess against the isospin-saturated core. As both of these structures have a very complex nature at the interface of coherent oscillations and pure particle-hole excitations and their microscopic texture is linked to the astrophysical r-process nucleosynthesis, they continuously attract active interest from both experiment and theory \cite{SavranAumannZilges2013}.

An accurate theory for the response of compound nuclei, which is required most often for applications, is extremely challenging. 
%However, a certain class of initial states, namely the compound states which are located in a (quasi)continuum, can be very well approximated by the thermal mean field as the excitation energy %in the high level density regime can be almost uniquely characterized by the temperature parameter. In this case, the desired response theory is a generalization of the ground-state response %theory to finite temperature. 
%However, 
The common practice 
%for finite-temperature microscopic approaches to nuclear response 
is to confine the framework by the simplest one-loop approximation, such as thermal random phase approximation (TRPA) \cite{RingRobledoEgidoEtAl1984,NiuPaarVretenarEtAl2009}, thermal quasiparticle RPA (TQRPA) \cite{RingRobledoEgidoEtAl1984,YuekselColoKhanEtAl2017}
or its  version with exact pairing \cite{HungDangHuong2017}, and one step further is represented by the continuum TQRPA \cite{LitvinovaKamerdzhievTselyaev2003,KhanVanGiaiGrasso2004,LitvinovaBelov2013}. 
A few extensions of T(Q)RPA 
%to 
including 
%important 
damping mechanisms have been formulated, for instance, in Ref. \cite{DukelskyRoepkeSchuck1998} as the finite-temperature self-consistent RPA, Ref. \cite{LacroixChomazAyik1998} in terms of the collision integral with incoherent two-particle two-hole (2p2h) excitations, Ref. \cite{Bortignon1986} as a coupling to collective surface vibrations in 
the framework of the nuclear field theory, and Refs. \cite{AdachiSchuck1989,Yannouleas1986} as the finite-temperature second RPA (SRPA) by the equation of motion method. The numerical implementations 
of the approaches beyond T(Q)RPA are very scarce and mainly incomplete. For instance, Ref. \cite{Bortignon1986} ignores coupling to non-collective modes and  Ref. \cite{LacroixChomazAyik1998} does not include collective effects, which leads to the controversial results: the latter model concluded on a rapid increase of the GDR's width with temperature 
%in $^{40}$Ca 
while the former found the GDR nearly unchanged. The theories of thermal shape fluctuations \cite{Kusnezov1998,OrmandBortignonBroglia1996} 
explain successfully the temperature dependence of the GDR's width found in experiments \cite{Gaardhoje1984a,GaardhojeEllegaardHerskindEtAl1986,Bracco1989,Ramakrishnan1996}, however, do not provide microscopic details of the nuclear strength functions. The evolution of the low-energy strength with temperature, which is crucial for astrophysical modeling, is barely addressed in these calculations.
%
%In order to meet the very high standards required for these applications, theoretical approaches to nuclear response must be highly accurate and at the same time based on fundamental %concepts of the nucleon-nucleon interaction. The latter provides an advanced predictive power and the former is of the utmost importance as the errors contained in nuclear strength functions %can propagate tremendously \cite{MumpowerSurmanMcLaughlinEtAl2016}. 

In this Letter, we present a new microscopic self-consistent approach to the nuclear response, which is free of the above mentioned limitations. We advance the approach developed previously in the zero-temperature framework of the relativistic nuclear field theory (RNFT) \cite{LitvinovaRing2006, LitvinovaRingTselyaev2007,LitvinovaRingTselyaev2008} to the finite-temperature case. The RNFT is based on the covariant energy density functional (CEDF) and extends the relativistic RPA (RRPA) beyond the one-loop approximation by taking into account the non-linear medium polarization effects in a parameter-free way. This theory performs very well when describing nuclear transitions from ground to excited states 
\cite{LitvinovaRingTselyaevEtAl2009,LitvinovaLoensLangankeEtAl2009,LitvinovaRingTselyaev2010,EndresLitvinovaSavranEtAl2010,
TamiiPoltoratskaNeumannEtAl2011,
MassarczykSchwengnerDoenauEtAl2012,LitvinovaRingTselyaev2013,SavranAumannZilges2013,LanzaVitturiLitvinovaEtAl2014,PoltoratskaFearickKrumbholzEtAl2014,Oezel-TashenovEndersLenskeEtAl2014,EgorovaLitvinova2016} and shows a similar potential for its finite-temperature generalization.  

%===============================================================================
% Formalism
{\it Formalism. \textemdash} To determine microscopic characteristics of a compound nucleus, we apply the finite-temperature mean-field
theory based on the variational principle of maximum entropy \cite{Sommermann1983} with the CEDF of Ref. \cite{VretenarAfanasjevLalazissisEtAl2005}.
%\bea
%E[\rho,\phi] = Tr[({\vec\alpha}{\vec p} + \beta m)\rho] + \sum\limits_m\Bigl\{Tr[(\beta\Gamma_m\phi_m)\rho] \mp \nonumber \\
%\mp \int d^3r \Bigl[\frac{1}{2} ({\vec\nabla}\phi_m)^2 + U(\phi_m)\Bigr]\Bigr\},
%\label{cedf}
%\eea
%where $\rho$ and $\phi$ are the nucleon densities and meson fields, respectively, and the meson potentials $U(\phi_m)$ are %treated within the non-linear sigma-meson model NL3 \cite{Lalazissis1997}. 
Here we consider nuclear systems without superfluid pairing, which include doubly-magic nuclei and other systems at temperatures above the critical temperature.
% when superfluidity vanishes. 
Thus, the eigenvalues of the nucleonic density matrix are the Fermi-Dirac occupation numbers
\be
%n_i = 
n_1(T) = n(\varepsilon_1, T) = \frac{1}{1 + e^{\varepsilon_1 /T} },
\ee
where the number index "$1$" runs over the complete set of the single-particle quantum numbers in the Dirac-Hartree basis, which includes
the single-particle energies $\varepsilon_1 = {\tilde\varepsilon_1} - \lambda$ measured from the chemical potential $\lambda$.

In order to compute the nuclear response to an external field of the particle-hole character, we have generalized the relativistic time blocking approximation of Ref. \cite{LitvinovaRingTselyaev2007} to the case of finite temperature by adopting the time blocking technique to the Matsubara temperature Green function formalism. The response function ${\cal R}_{14,23}(\omega,T)$ in the frequency ($\omega$) domain, thus, reads:
\bea
{\cal R}_{14,23}(\omega,T) &=& \tilde{\cal R}_{14,23}(\omega,T) + \sum\limits_{1'2'3'4'}\tilde{\cal R}_{12',21'}(\omega,T)\times  \nonumber \\ 
&\times& {\cal W}_{1'4',2'3'}(\omega,T) 
{\cal R}_{3'4,4'3}(\omega,T),
\label{bse3}
\eea
where
$\tilde{\cal R}(\omega,T)$ is the uncorrelated particle-hole propagator
\begin{equation}
\tilde{\cal R}_{14,23}(\omega,T) = \frac{n_{21}(T)\delta_{13}\delta_{24}}{\omega - \varepsilon_{1} + \varepsilon_{2} }, 
\label{freeresp}
\end{equation}
$n_{21}(T) = n_2(T )- n_1(T)$, and the interaction kernel transforms to the $\omega$-domain as:
\be
{\cal W}_{{1}{4},{2}{3}}(\omega,T)=\tilde{V}%
_{{1}{4},{2}{3}}(T)
%\nonumber \\
+ \Phi_{{1}{4},{2}%
{3}}(\omega,T)-\Phi_{{1}{4},{2}{3}}(0,T).
\label{W-omega-tba}%
\ee
The particle-vibration coupling (PVC) amplitude $\Phi(\omega,T)$ is given by:
\begin{eqnarray}%
\Phi_{{1}{4},{2}{3}}^{(ph)}(\omega,T)  = \frac{1}{n_{43}(T) } 
\sum\limits_{56\mu} \sum\limits_{\eta_{\mu}=\pm1}
\eta_{\mu} \zeta^{\mu\eta_{\mu}}_{12,56}  \zeta^{\mu\eta_{\mu}\ast}_{34,56}\times \nonumber \\
\times\frac{ \bigl(N(\eta_{\mu}\Omega_{\mu}) + n_6(T)\bigr)\bigl(n(\varepsilon_{6}-\eta_{\mu}\Omega_{\mu},T) - n_5(T)\bigr)}{\omega-\varepsilon_{5}+\varepsilon_{6}-\eta_{\mu}\Omega_{\mu} }, 
\nonumber\\
\label{phiph}%
%\end{align}
\end{eqnarray} 
where we denote the phonon vertex matrices $\zeta^{\mu\eta_{\mu}}$ as:
\be
\zeta^{\mu\eta_{\mu}}_{12,56} = \delta_{15}\gamma^{\eta_{\mu}}_{\mu;62} - \gamma^{\eta_{\mu}}_{\mu;15}\delta_{62},
\label{zetas}
\ee
with the matrix elements of the particle-phonon coupling vertices %$\gamma_{\mu;13}^{\eta_{\mu}}$ 
%\begin{equation}
$\gamma_{\mu;13}^{\eta_{\mu}} = \delta_{\eta_{\mu},+1}\gamma_{\mu;13} + \delta_{\eta_{\mu},-1}\gamma_{\mu;31}^{\ast}$,
%\end{equation}
and the phonon frequencies as $\Omega_{\mu}$. The index $\mu$ numerates the phonon quantum numbers and, as in Ref. \cite{LitvinovaRingTselyaev2007}, the vertices $\gamma_{\mu;13}$ are extracted from the solution of Eq. (\ref{bse3}) without the frequency-dependent interaction, which corresponds to  the finite-temperature relativistic random phase approximation (FT-RRPA). The bosonic occupation factors
$N(\Omega) = 1/ (e^{\Omega /T}-1)$ in Eq. (\ref{phiph}) emerge as additional phonon characteristics.
%naturally during the Fourier transformation of the interaction kernel to the manifold of its frequency variables. 
The hole-particle $hp$-components of the amplitude $\Phi(\omega,T)$ are found analogously. At finite temperatures we 
%imply the following convention: 
consider a pair of states $\{12\}$ 
%is regarded 
as a $ph$-pair if $n_{21}(T) > 0$ and as an $hp$-pair if $n_{21}(T) < 0$, while pairs with $n_{21}(T) = 0$ are eliminated by Eq. (\ref{freeresp}). In Eq. (\ref{W-omega-tba}) this amplitude is corrected by the subtraction of itself at zero frequency, in order to avoid double counting of the particle-phonon coupling effects, which are implicitly included in the mean field, and provides a good convergence of the amplitude $\Phi(\omega,T)$ with respect to the phonon space \cite{Tselyaev2007,Tselyaev2013}.  The latter feature is widely utilized in applications  as it  allows a truncation of the phonon model space by the phonon energy $\sim$15-20 MeV in calculations for medium-heavy nuclei. At T=0, $\Phi(\omega)$ is known to induce damping effects which lead to an additional fragmentation and broadening of the strength functions
%, as compared to the case of keeping only the instantaneous interaction $\tilde V$ in Eq. (\ref{W-omega-tba}) 
\cite{LitvinovaRingTselyaev2007,LitvinovaRingTselyaev2008,MarketinLitvinovaVretenarEtAl2012,LitvinovaBrownFangEtAl2014,RobinLitvinova2016}. As it is shown below, at finite temperature this amplitude plays a similar role. 
%It is also easy to verify that 
%the T$\to$ 0 limit of the amplitude $\Phi(\omega,T)$ matches its analog in the zero-temperature response theory %\cite{LitvinovaRingTselyaev2007}.

Self-consistent calculations within the approach of Eqs. (\ref{bse3}-\ref{zetas}) named finite-temperature relativistic time blocking approximation (FT-RTBA) were performed in a wide range of temperatures for the dipole response of the three nuclei, $^{120,132}$Sn and $^{48}$Ca, for which we have obtained a very good description of data at T=0 \cite{LitvinovaRingTselyaev2008,LitvinovaRingTselyaevEtAl2009,EgorovaLitvinova2016}, and for an unstable $^{100}$Sn nucleus with equal numbers of protons and neutrons (N=Z).  For each temperature value, we solved first the self-consistent relativistic mean field (RMF) problem with NL3 forces \cite{Lalazissis1997}, which determined the meson and nucleon fields and the single-nucleon Dirac-Hartree basis. In this basis, the FT-RRPA problem with the same meson-exchange interaction was solved to obtain the vertices and frequencies of the phonon modes. 
%In the present application we include only usual isoscalar phonons which are proven to play the leading role in the %fragmentation of single-particle and collective states, although contributions from the isovector phonons can be also sizable %\cite{Litvinova2016}. 
The phonon space was truncated on  angular momenta, frequencies and reduced transition probabilities using the same criteria as in earlier calculations for T=0. We included isoscalar phonon modes with natural parities and multipolarities $2\leq J_{\mu} \leq$ 6, frequencies $\Omega_{\mu}\leq \Omega_c$ ($\Omega_c = 15$ MeV for $^{120,132}$Sn and $\Omega_c = 20$ MeV for $^{48}$Ca and $^{100}$Sn),  and the reduced transition probabilities exceeding 5\% of the maximal one for each multipolarity. Keeping these criteria, we have included coupling to both collective and non-collective states while the total number of phonon modes grows by up to a factor of ten at T$\sim$ 5-6 MeV, compared to T=0. The selected phonon modes, together with the single-particle output of the thermal RMF, determine the particle-phonon coupling amplitude $\Phi(\omega,T)$ and, thus, the response function of Eq. (\ref{bse3}). Finally, 
the microscopic strength function ${\tilde S}(E,T)$ is defined as a response of a compound nucleus to an external field $V^0$ \cite{RingRobledoEgidoEtAl1984}: ${\tilde S}(E,T) = S(E,T)/(1-e^{-E/T})$, where
\be
S(E,T) = -\frac{1}{\pi}\lim\limits_{\Delta\rightarrow+0}Im\
\langle V^{0\dagger} {\cal R}(E + i\Delta,T)V^0 \rangle
\label{strf}%
\ee
%
%\be
%P_{LS}^{J} = \sum\limits_{i=1}^{A} r^L(i)[\sigma_S (i)\otimes Y_L(i)]^{J}\tau_{0}(i)
%P_{L\pm}^{\lambda} = \sum\limits_{i=1}^{A} r^L(i)[\sigma (i)\otimes Y_L(i)]^{\lambda}\tau_{\pm}(i)
%\label{extfield}
%\ee
%\begin{equation}
 %\ \ 
%\end{equation}
with $V^0$ being a dipole operator and with a finite value of the imaginary part of the energy variable $\Delta$, which accounts for missing effects of higher-order configurations and the continuum \cite{LitvinovaRingTselyaev2008}.
%usually in the calculations it has a finite value which helps to include some missing mechanisms in an effective way. The missing mechanisms here can be particle escape to the continuum and configurations more complex than those included explicitly. 
%\begin{figure}[ptb]
%\vspace{-2cm}
%\begin{center}
%\includegraphics[scale=0.22]{E_vsT_k.pdf}
%\end{center}
%\vspace{-0.5cm}
%\caption{Energy of the compound nucleus as a function of temperature for  $^{48}$Ca and $^{132}$Sn: RMF (blue symbols) %and parabolic fits (red curves).}
%\label{EvsT}
%\end{figure}
%
\begin{figure}[ptb]
\vspace{0.2cm}
\begin{center}
\includegraphics[scale=0.30]{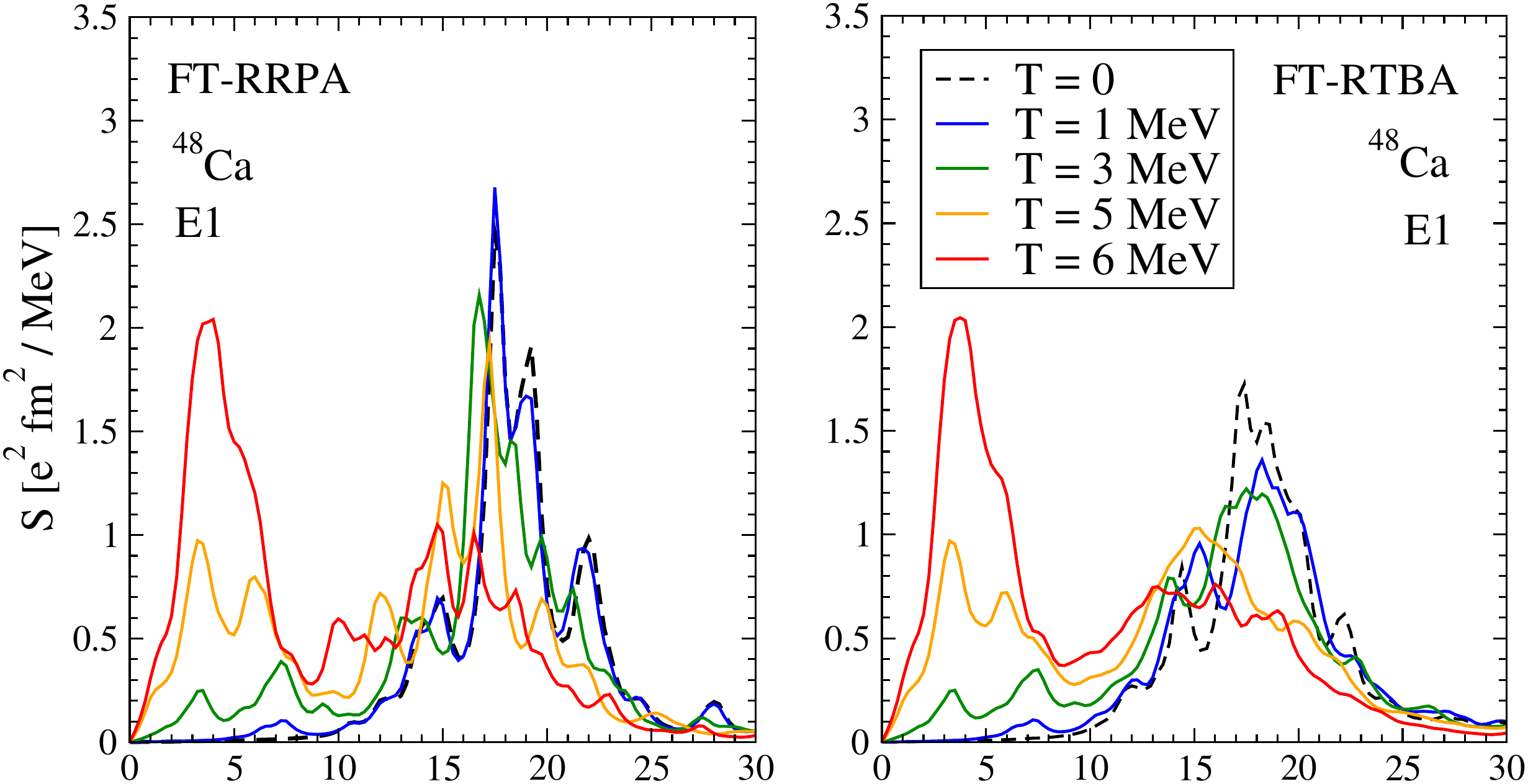}
%\vspace{-2cm}
\includegraphics[scale=0.30]{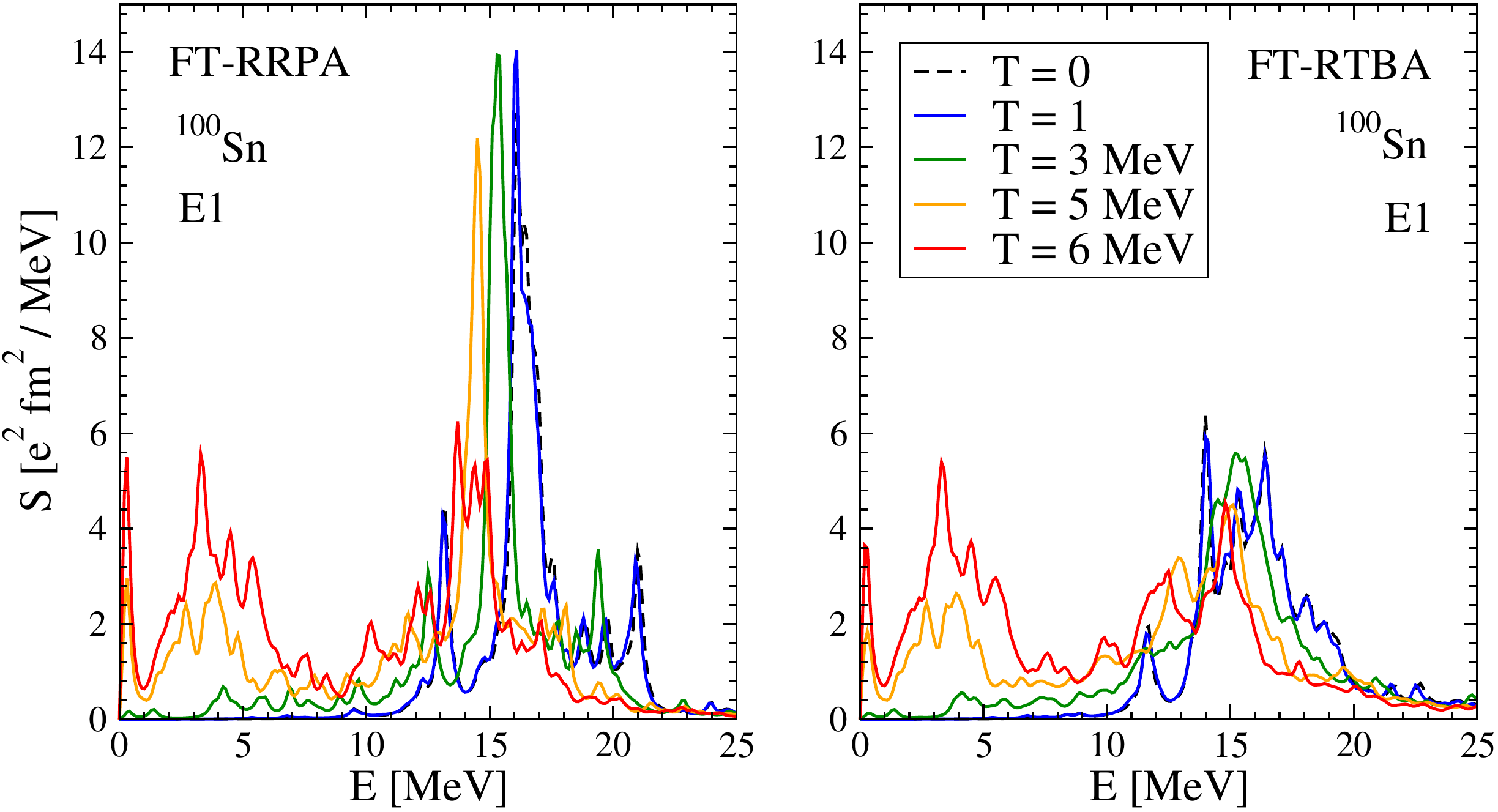}
\end{center}
\vspace{-0.5cm}
\caption{Electric dipole strength distribution in $^{48}$Ca and $^{100}$Sn calculated within FT-RRPA (left panels) and FT-RTBA (right panels) at various temperatures. The values of the imaginary part of the energy variable $\Delta$ = 500 keV and  $\Delta$ = 200 keV were adopted for $^{48}$Ca and $^{100}$Sn, respectively.
}
\label{casn}
\end{figure}
\begin{figure}[ptb]
%\vspace{0.2cm}
\begin{center}
\includegraphics[scale=0.30]{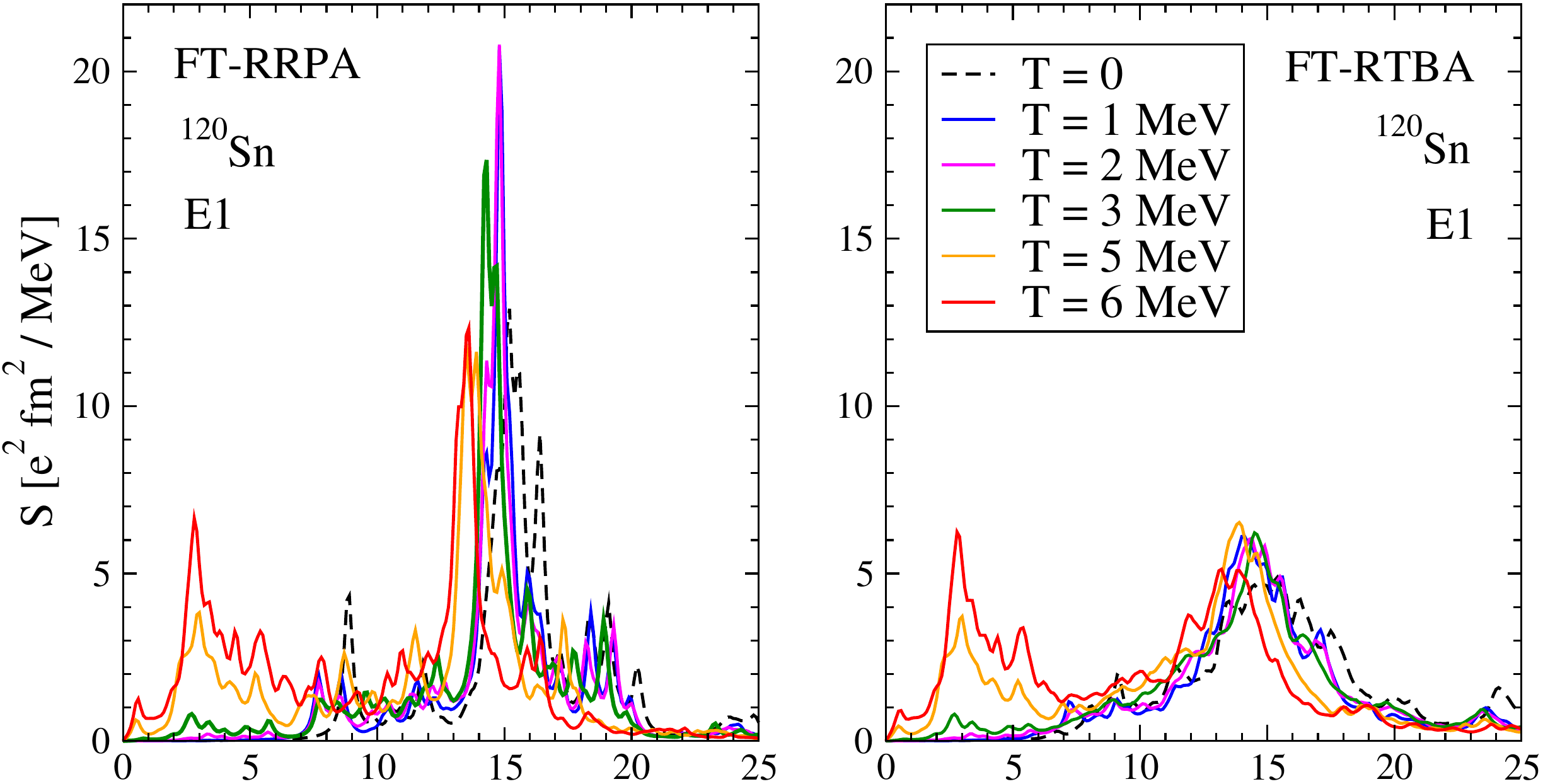}
%\vspace{2cm}
\includegraphics[scale=0.30]{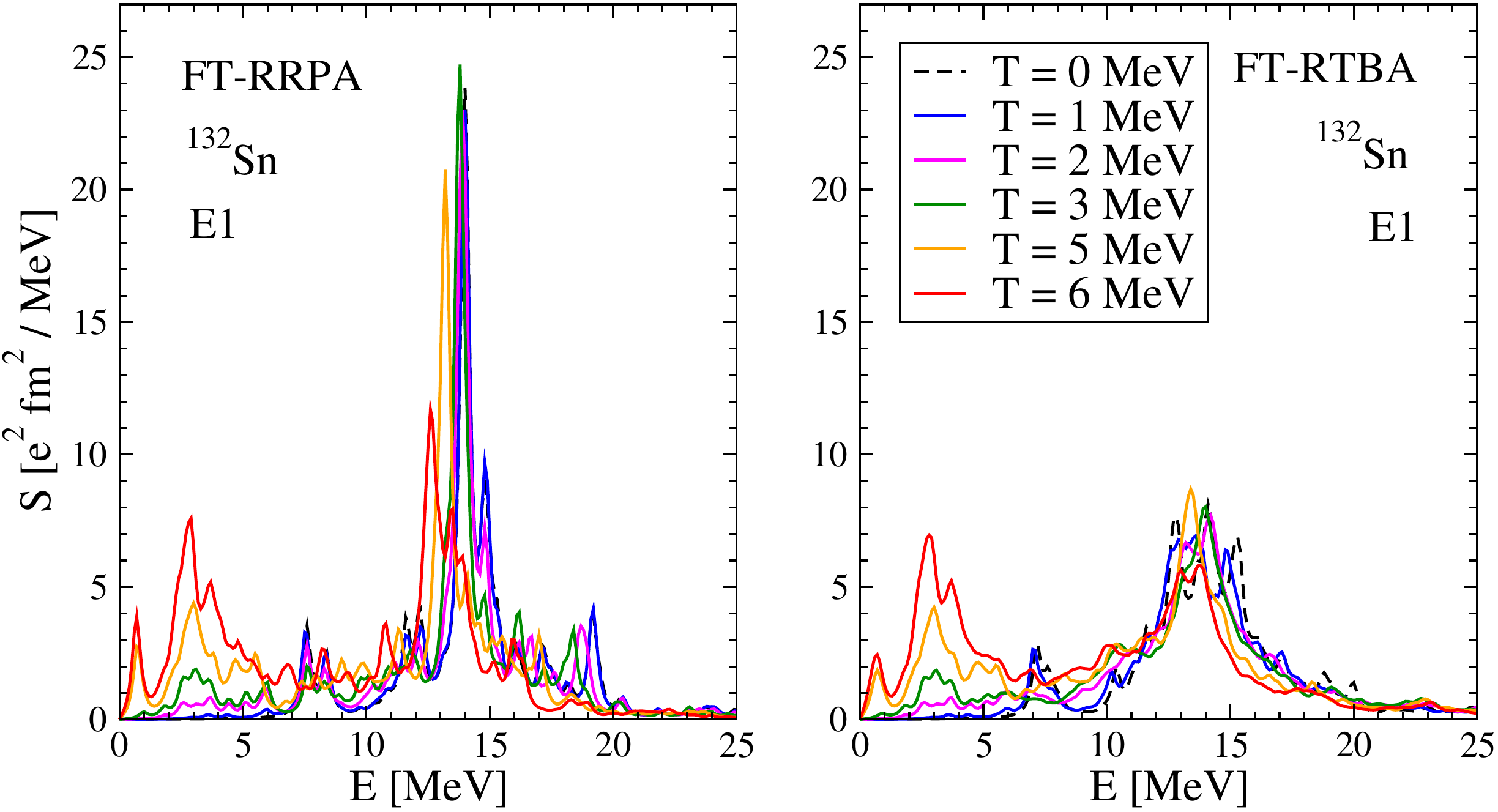}
\end{center}
\vspace{-0.5cm}
\caption{Same as in Fig. \ref{casn} but for $^{120,132}$Sn with $\Delta$ = 200 keV.}
\label{snsn}
\end{figure}
%
%\begin{figure}[ptb]
%\vspace{-1.5cm}
%\begin{center}
%\includegraphics[scale=0.35]{120sn_e1_200.eps}
%\end{center}
%\vspace{-1.5cm}
%\caption{Electric dipole strength distribution in $^{48}$Ca and $^{120}$Sn calculated within FT-RRPA (left panels) and FT-%RTBA (right panels) at different temperatures.}
%\caption{Same as in Fig. \ref{48ca} but for $^{120}$Sn.}
%\label{120sn}
%\end{figure}

{\it Results. \textemdash}The results for the dipole strength distributions $S(E,T)$ in a medium-light $^{48}$Ca nucleus and in  medium-heavy $^{100,120,132}$Sn nuclei at various temperatures are displayed in Figs. \ref{casn} and \ref{snsn}, where we compare the evolution of the strength within FT-RRPA 
%(left panels) 
and FT-RTBA.
%(right panels). 
In $^{48}$Ca we observe the two major effects with the temperature increase: a continuous broadening and a quenching of the GDR and an enhancement of the low-energy strength associated with the 
%pygmy dipole resonance
PDR. This evolution is accompanied by a slow movement of the 
%GDR as a whole 
entire distribution towards the lower energy.
%, so that the energy weighted sum rule of the entire dipole spectrum is preserved within a few percent.
At 
%the temperature of 
T$\approx$ 5-6 MeV the low-energy part begins to dominate for all multipoles, which, in turn, reinforces the damping of the GDR.  
%In turn, at T=6 MeV GDR becomes very fragmented while the low-energy peak grows tremendously. 
These  temperatures are consistent with the limit of existence of the GDR established in earlier works \cite{Santonocito2006}.
The obtained evolution of the dipole strength with the temperature increase is caused by 
%the interplay of several factors: 
(i) a slow change of the self-consistent mean field and of the single-particle orbits of the compound nucleus, (ii) the increasing diffuseness of the Fermi surface which enhances the amount of the low-energy particle-hole configurations and reinforces the Landau damping, 
%which is seen already in FT-RRPA, 
(iii) a reduction of the leading contribution to the particle-phonon coupling amplitude 
%$\Phi(\omega,T)$ 
(\ref{phiph}) associated with $\eta_{\mu} =1$, which is the only contribution at T=0, and an increasing role of the new terms with $\eta_{\mu} =-1$.
%, which generate new poles and give rise to further fragmentation of the strength in FT-RTBA. 

%Notice that in Figs. \ref{casn}, \ref{snsn} we plot the microscopic strength functions $S(\omega,T)$ without the exponential %factor, which is present in ${\tilde S}(\omega,T)$ due to the detailed balance, in order to see the details of the nuclear %response at very low transition energy $E$.
%This factor does not affect the GDR region at all temperatures under study, however, at moderate to high temperatures it %enhances the low-energy peak by up to a factor of two. This factor also affects the zero-energy limit of the strength %distribution bringing it to a finite value and will be studied in detail in future work. 

The trends are similar in the medium-heavy tin nuclei. $^{120}$Sn is superfluid below the critical temperature $T_c\approx 0.66$ MeV in our framework and $^{100,132}$Sn have closed shells in both proton and neutron subsystems. 
Thus,
%This means that 
the GDR's width in $^{120}$Sn decreases in the temperature interval $0\leq T \leq T_c$ and begins to increase at $T>T_c$, while the GDR's widths in $^{100,132}$Sn grow in the entire temperature range. 
%
%Fragmentation
Comparing right and left panels of Figs. \ref{casn}, \ref{snsn} one can notice that the fragmentation 
effect of the PVC mechanism 
is relatively weak for the excited states emerging with the temperature increase at low energy. These states are absent at T=0 and formed solely by the transitions within thermally unblocked pairs. It turns out that for such pairs of states the matrix elements of the PVC amplitude $\Phi(\omega)$ are strongly suppressed by the factor in the numerator of  Eq. (\ref{phiph}) $n(\varepsilon_{6}-\eta_{\mu}\Omega_{\mu},T) - n_5(T)$. 
%The smallness of this factor for the thermally unblocked pairs is not compensated by the denominator $1/n_{43}(T)$ which is balanced by the numerator of the free propagator. 
Thus, the lack of fragmentation of the thermally unblocked states is, probably, due to some limitations of our approach: the PVC amplitude $\Phi(\omega)$ includes only the so-called resonant part and does not include the ground state correlations (GSC) induced by the PVC. The structure of these GSC/PVC terms at T=0 \cite{KamerdzhievTertychnyiTselyaev1997} points out that they may enforce the fragmentation effect on the thermally unblocked transitions. The inclusion of these effects for  T$>$0 will be considered elsewhere. 

%We also conclude that the limiting temperature for the nuclei $^{120,132}$Sn is by $\sim$1 MeV higher than the one for  
%$^{48}$Ca.
%
\begin{figure}[ptb]
%\vspace{-0.5cm}
\begin{center}
\includegraphics[scale=0.35]{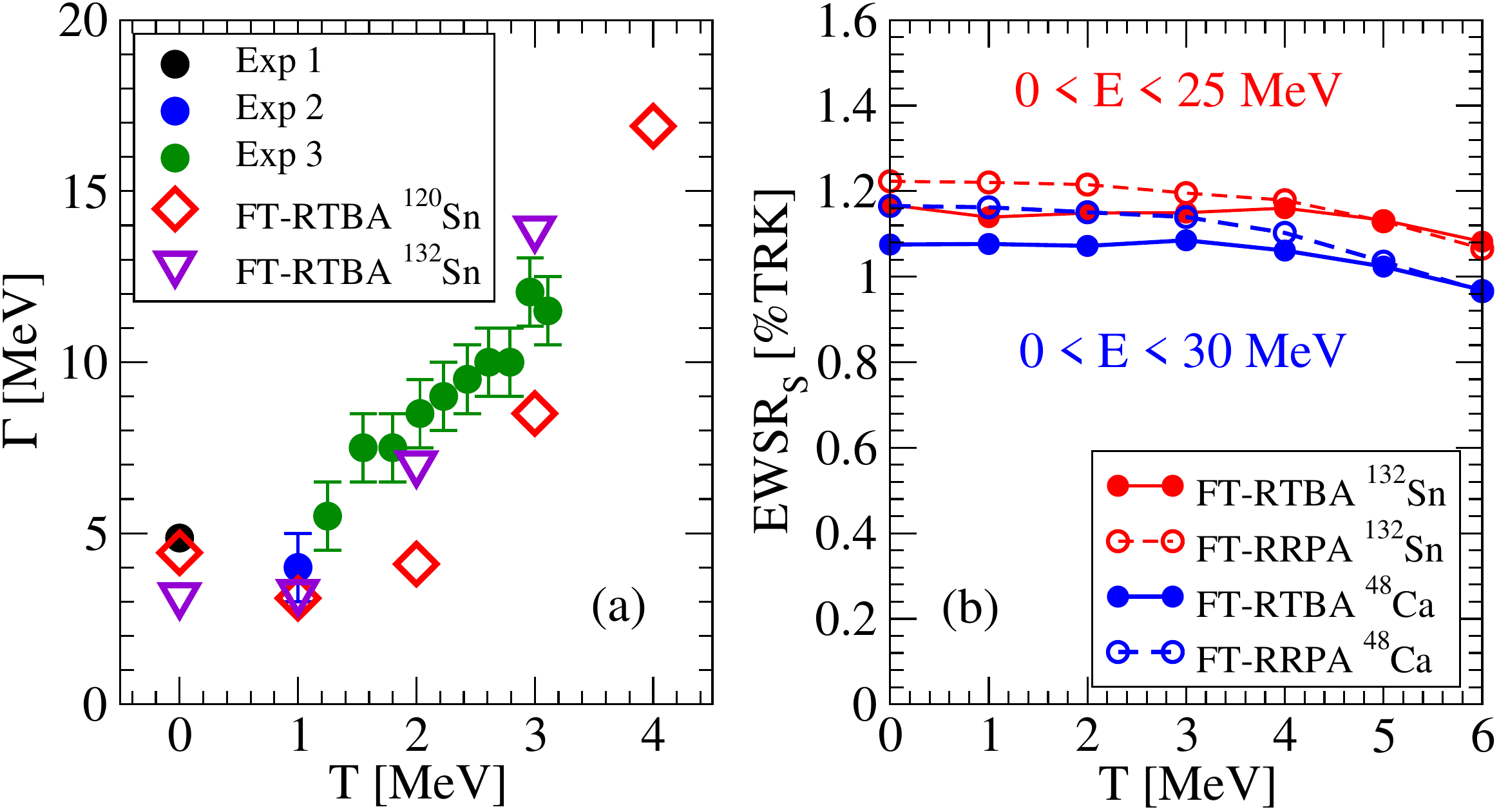}
\end{center}
\vspace{-0.5cm}
\caption{Left panel: Width of the giant dipole resonance in $^{120,132}$Sn as a function of temperature. The experimental values from Refs. \cite{Fultz1969,Heckman2003,Ramakrishnan1996} are shown for $^{120}$Sn. Right panel: The energy-weighted sum rule (EWSR) for $^{48}$Ca and $^{132}$Sn with respect to the TRK sum rule.}
\label{GDR_Gamma}
\end{figure}
%
%. Widths and sum rule
%
The temperature dependence of the GDR's width $\Gamma(T)$ obtained in FT-RTBA is shown in Fig. \ref{GDR_Gamma} (a) for  $^{132}$Sn and $^{120}$Sn together with experimental data 
%which are 
available 
%only 
for $^{120}$Sn. The theoretical widths at T=0 are taken from our previous calculations \cite{LitvinovaRingTselyaev2007,LitvinovaRingTselyaev2008}.  
The nucleus $^{120}$Sn is superfluid below T $\sim$ 0.66 MeV and the $^{132}$Sn one is non-superfluid. 
%even at T=0. 
Because of the phase transition in $^{120}$Sn at T $\approx$ 0.66 MeV, $\Gamma(T)$ has a smaller value at T=1 MeV than at T=0 as disappearance of the superfluid pairing reduces the width.
% compared to the case of superfluidity. 
Thermal effects are not yet seen at T=1 MeV 
%and the reason can be understood by looking at the 
because of specific
shell structure of $^{120}$Sn. When the temperature starts to increase, the protons which form the closed shell have the next available orbitals only in the next major shell, and T=1 MeV temperature is not sufficient to promote the protons over the shell gap with a noticeable occupancy. In the neutron subsystem, the lowest available orbit is the intruder $1h_{11/2}$ state where particles get promoted relatively easily, but after this orbit there is another shell gap. As a consequence, at T=1 MeV there is still no room for a noticeable thermal unblocking. Thus, our result can explain the unexpectedly small GDR's width at T=1 MeV reported in Ref. \cite{Heckman2003}, in contrast to the thermal shape fluctuation calculations.

In $^{132}$Sn we observe a relatively fast increase of $\Gamma(T)$ after T=1 MeV and no increase below this temperature because of the presence of the shell gaps in both proton and neutron subsystems.
%, in analogy to the case of proton subsystem in $^{120}$Sn discussed above. 
After T=1 MeV in $^{132}$Sn and T=2 MeV in $^{120}$Sn we obtain a fast increase of $\Gamma(T)$ because of (i) the formation of the low-energy shoulder and (ii) a slow increase of the fragmentation of the high-energy peak which becomes faster at high temperatures when the low-energy phonons develop the new sort of collectivity. As $^{132}$Sn is more neutron-rich than $^{120}$Sn, the  respective strength in the low-energy shoulder of $^{132}$Sn is larger, which leads to a larger overall width in $^{132}$Sn at temperatures above 1 MeV. Note that  $\Gamma(T)$ values for T$>$3 MeV in $^{132}$Sn and at T$>$4 MeV in $^{120}$Sn are not presented because the standard procedure based on the Lorentzian fit of the microscopic strength distribution fails in recognizing the distribution as a single peak structure.

For $^{120}$Sn
%, where experimental information on $\Gamma(T)$ is available up to T$\approx$3 MeV, 
we find a good agreement with data for T=0, T=1 MeV and a reasonable agreement for T=3 MeV. The discrepancy at T=2 MeV can be due to the fact that in the experimental studies in this temperature range compound nuclei acquire non-zero angular momenta up to 60 $\hbar$. This, in turn, increases the total width of the GDR \cite{Mattiuzzi1997}. We do not have such effects included in the presented approach, but they can be considered in the future work. It is also established experimentally that at temperatures around 3  MeV the angular momentum dependence of the GDR's width in tin nuclei is saturated \cite{Santonocito2006}, which is consistent with our finding that at T=3 MeV our calculated $\Gamma(T)$ in $^{120}$Sn is getting much closer to the data.
%The overall agreement of FT-RTBA calculations with data is found very reasonable except for the range of temperatures 1 %MeV$\leq T \leq$ 3 MeV, possibly
%due to high angular momenta and deformation effects which are not included in the present calculations. Our results are %consistent with those of a microscopic approach of Ref. \cite{Bortignon1986}, which are available for the GDR energy %region at $T\leq$ 3 MeV,
%while in the entire range of temperatures under study $\Gamma(T)$ shows a nearly quadratic dependence %agreeing with the Fermi liquid theory \cite{Landau1957}.

%Due to the various factors discussed above, in FT-RTBA the spreading effects accrue rather slowly until reaching the range %of the limiting temperature, while the Landau damping intensifies tremendously. As a result, at high temperatures of 5-6 %MeV the low-energy modes acquire a strong coherence, thereby enhancing the particle-phonon coupling and, hence, the %spreading of the GDR to the limits of its existence when it can no longer be recognized as a single-peak structure. 
%
%. Sum rule
%
Fig. \ref{GDR_Gamma} (b) illustrates the evolution of the dipole energy-weighted some rule (EWSR), which is given in \% with respect to the Thomas-Reiche-Kuhn (TRK) sum rule, in $^{48}$Ca and  $^{132}$Sn. As shown in Ref. \cite{Barranco1985a}, the EWSR at T$>$0 can be calculated in full analogy with the case of T=0. Since the meson-exchange interaction is velocity-dependent, already in RRPA and its superfluid version at T=0 we observe up to  40\% enhancement of the 
%Thomas-Reiche-Kuhn (TRK) 
TRK sum rule within the experimentally studied energy regions \cite{LitvinovaRingTselyaev2007,LitvinovaRingTselyaev2008}, in agreement with data. In the resonant time blocking approximation without the GSC/PVC the EWSR should have exactly the same value as in RPA, however, the subtraction procedure causes a little violation of it \cite{Tselyaev2007,LitvinovaTselyaev2007}. In the RTBA vs RRPA at T=0 we usually find a few percent less in finite energy intervals below 25-30 MeV, but this difference decreases if we take larger intervals
%. This is due to the fact that 
since in RTBA the strength distributions are more spread and have heavier high-energy tails.
% which, if cut, leaves more strength outside the finite interval. 
A similar situation takes place at T$>$0. Fig. \ref{GDR_Gamma} (b) shows a slow decrease of the EWSR with temperature because the entire resonance moves down in energy while FT-RRPA and FT-RTBA values practically meet at T=6 MeV when their high-energy tails become less important.

%  Transition densities
%
\begin{figure}[ptb]
%\vspace{-0.5cm}
\begin{center}
\includegraphics[scale=0.35]{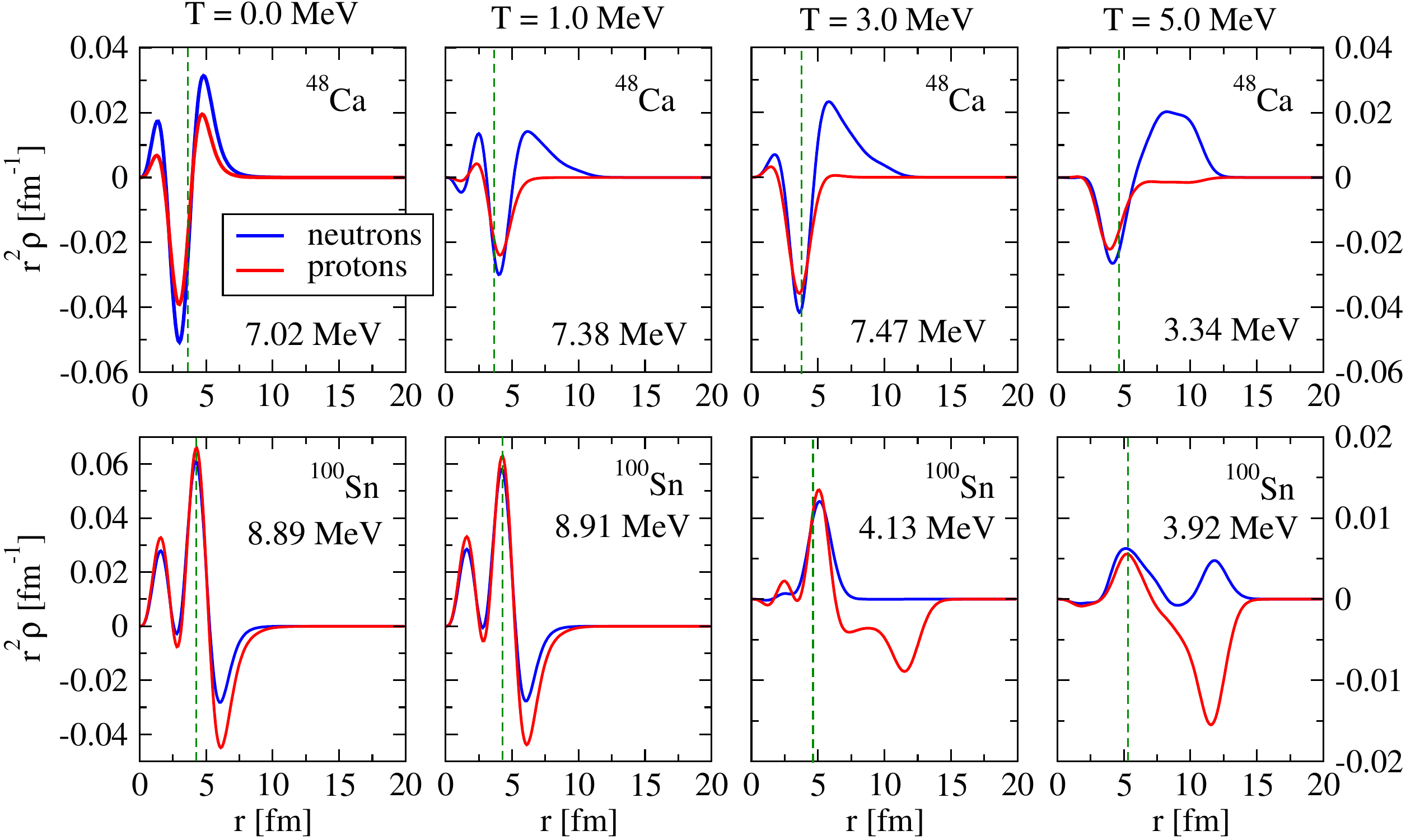}
\end{center}
\vspace{-0.5cm}
\caption{Temperature evolution of the proton and neutron transition densities for the most prominent peaks below 10 MeV in $^{48}$Ca and $^{100}$Sn within FT-RTBA. The dashed vertical lines indicate the rms nuclear radius.}
\label{TrDen}
\end{figure}
Figure \ref{TrDen} displays the temperature evolution of transition densities in the neutron and proton subsystems of the neutron-rich $^{48}$Ca in comparison to those of the N=Z neutron-deficient $^{100}$Sn, for their strongest peaks below 10 MeV. This energy region is typically associated with the PDR. One can see that at all temperatures 48-Ca shows a usual pattern of the in-phase oscillation of protons and neutrons inside the nucleus while the neutron oscillation dominates in the outside area. The radial transition densities become more extended with the temperature growth, but qualitatively their behavior remains similar. An analogous situation occurs in $^{100}$Sn, but with the proton dominance outside. The entire PDR slowly moves toward lower energies with the temperature growth in both nuclei. Large spatial extensions of the radial transition density distributions for the leading low-energy peaks at high temperatures may point out to approaching the limits of existence of the considered nuclear systems.

%===============================================================================
%\section{Summary}
%\label{summary}
{\it Summary. \textemdash} We present an advanced microscopic self-consistent many-body approach to the finite-temperature nuclear response, which includes spreading effects in addition to Landau damping. In this framework we investigated the evolution of the dipole strength distribution in medium-mass nuclei in a wide range of temperatures.  The obtained results are consistent with the existing experimental data on the GDR's width and with the Landau theory \cite{Landau1957}, explain the critical phenomenon of the disappearance of the GDR at high temperatures and predict the evolution of the low-energy dipole strength with temperature taking into account spreading effects.

The analytical method presented in this work is quite general and can be widely applied to investigate the response of strongly-correlated systems  at finite temperature. 
The presented numerical implementation of the developed approach opens a way to accurate systematic studies of excitations and de-excitations of compound nuclei in a wide energy range. Both giant resonances and soft modes of various multipolarities in medium-heavy nuclei are of primary interest in this context because of their direct 
implications for r-process nucleosynthesis. Such systematic studies will be addressed by future efforts.

%===============================================================================
%\section*{Acknowledgements}
The authors greatly appreciate discussions with P. Schuck and D. Halderson. This work is partly supported by US-NSF grant PHY-1404343 and by NSF Career grant PHY-1654379.
%
%===============================================================================

\bibliography{Bibliography_Jul2018}
\end{document}